\documentclass[trackchanges,twocolumn,resetfootnote]{aastex701}
\usepackage{amsmath}
\usepackage[export]{adjustbox}

\begin{document}

\title{The formation and evolution of dust in the colliding-wind binary Apep revealed by JWST}

\author[orcid=0000-0002-2106-0403, sname='Han']{Yinuo~Han}
\affiliation{Division of Geological and Planetary Sciences, California Institute of Technology, 1200 E. California Blvd., Pasadena, CA 91125, USA}
\email[show]{yinuo@caltech.edu}

\author[orcid=0009-0006-7054-0880,sname='White']{Ryan~M.~T.~White}
\affiliation{School of Mathematical and Physical Sciences, 12 Wally's Walk, Macquarie University, Sydney, 2113, NSW, Australia}
\affiliation{School of Mathematics and Physics, University of Queensland, Brisbane, 4072, QLD, Australia}
\email{ryan.white@mq.edu.au}

\author[0000-0002-7167-1819, sname='Callingham']{Joseph~R.~Callingham}
\affiliation{ASTRON, Netherlands Institute for Radio Astronomy, Oude Hoogeveensedijk 4, Dwingeloo, 7991 PD, The Netherlands}
\affiliation{Anton Pannekoek Institute for Astronomy, University of Amsterdam, Science Park 904, 1098\,XH, Amsterdam, The Netherlands}
\email{jcal@strw.leidenuniv.nl}

\author[0000-0003-0778-0321, sname='Lau']{Ryan~M.~Lau}
\affiliation{NSF NOIRLab, 950 N. Cherry Ave. Tucson, AZ 85719, USA}
\affiliation{IPAC, Mail Code 100-22, Caltech, 1200 E. California Blvd., Pasadena, CA 91125, USA}
\email{ryanlau@ipac.caltech.edu}

\author[orcid=0000-0003-2595-9114,sname='Pope']{Benjamin~J.~S.~Pope} 
\affiliation{School of Mathematical and Physical Sciences, 12 Wally's Walk, Macquarie University, Sydney, 2113, NSW, Australia}
\affiliation{School of Mathematics and Physics, University of Queensland, Brisbane, 4072, QLD, Australia}
\email{benjamin.pope@mq.edu.au}

\author[0000-0002-2806-9339, sname='Richardson']{Noel~D.~Richardson}
\affiliation{Department of Physics and Astronomy, Embry-Riddle Aeronautical University, 3700 Willow Creek Road, Prescott, AZ 86301, USA}
\email{RICHAN11@erau.edu}

\author[orcid=0000-0001-7026-6291, sname='Tuthill']{Peter~G.~Tuthill}
\affiliation{Sydney Institute for Astronomy, School of Physics, The University of Sydney, Sydney, 2006, NSW, Australia}
\email{peter.tuthill@sydney.edu.au}

\begin{abstract}
Carbon-rich Wolf-Rayet (WR) stars are significant contributors of carbonaceous dust to the galactic environment, however the mechanisms and conditions for formation and subsequent evolution of dust around these stars remain open questions. Here we present JWST observations of the WR+WR colliding-wind binary Apep which reveal an intricate series of nested concentric dust shells that are abundant in detailed substructure. The striking regularity in these substructures between successive shells suggests an exactly repeating formation mechanism combined with a highly stable outflow that maintains a consistent morphology even after reaching 0.6\,pc (assuming a distance of 2.4\,kpc) into the interstellar medium. The concentric dust shells show subtle deviations from spherical outflow, which could reflect orbital modulation 
along the eccentric binary orbit or non-sphericity in the stellar wind. Tracking the evolution of dust across the multi-tiered structure, we measure the dust temperature evolution that can broadly be described assuming an amorphous carbon composition in radiative thermal equilibrium with the central stars. The temperature profile and orbital period place new distance constraints that support Apep being at a greater distance than previously estimated, reducing the line-of-sight and sky-plane wind speed discrepancy previously thought to characterise the system. 
\end{abstract}



\section{Introduction}
Evolved stars are significant contributors to cosmic dust. Carbon-rich Wolf-Rayet stars, the descendants of the most massive O stars ($\gtrsim25 \, \mathrm{M}_\odot$, \citealp{Crowther2007}), are a major producer of carbon dust in the galactic environment \citep{Lau2020a, Lau2021}. These stars embody the final phase of evolution before being destroyed in core-collapse supernovae. They are characterised by rapid mass loss ($\sim10^{-5} \, \mathrm{M}_\odot \, \mathrm{yr}^{-1}$) driven by extreme luminosities ($\sim10^6 \, \mathrm{L}_\odot$), with a small fraction ($\lesssim$1\%) of the outflow mass nucleating into dust. 

While significant uncertainty surrounds the exact mechanism of dust formation around WR stars, resolved imaging of dust structures around WR stars found in colliding-wind binaries have provided clues on the dust formation conditions in such systems (e.g., \citealp{Tuthill1999b, Tuthill1999, Tuthill2006b, Lau2020b, Callingham2019, Han2020}). These observations provide a link between the binary orbit, the physical properties of the stellar winds and the production of dust. For the majority of dusty systems, where the binary consists of a carbon-type WR star and a massive (O- or B-type) companion, the collision of the component stellar winds creates a dense, carbon-rich environment that favours dust nucleation \citep{White2024}. The WR component, which displays the stronger wind momentum, forces the conical shock interface to bow towards the companion. Dust forms and propagates along this shock surface, streaming radially away from the star, while at the same time orbital motion azimuthally offsets the launch direction of newer dust continuously.
The net outcome of these two motions, a radially inflating wind structure injected at a rotating flow insertion point, results in a large-scale dust geometry that follows a spiral geometry. 

Apep (2XMM J160050.7–514245) is the most recently discovered colliding-wind WR binary to exhibit such a geometry \citep{Callingham2019}. Whereas other WR binaries typically involve an O- or B-type companion, Apep consists of two classical WR stars \citep{Callingham2020}. The WN4-6b star is thought to launch the stronger wind at 3500\,km\,s$^{-1}$ \citep{Callingham2020}, compared to  2100\,km\,s$^{-1}$ for the carbon-rich WC8 stellar wind. 
The combined action of these colliding winds was found to sculpt a large ($12^{\prime\prime}$) and intricate dust shell visible in the mid-infrared. Notably, a discrepancy (by a factor of $\sim$4) has been reported between the aforementioned line-of-sight wind speed measured with spectroscopy and the sky-plane expansion speed of its circumstellar dust \citep{Callingham2019, Han2020}. 
This finding is puzzling both from the perspective of the basic kinematic physics understood for these systems, and also from studies of other colliding wind spirals \citep{Tuthill1999, Tuthill2008, Lau2020b}, both of which lead to an expectation that the proper motions of the dust expansion should be consistent with the observed spectroscopic wind speed. This has led to the suggestion that the wind velocity field of one WR component may be asymmetric, possibly due to rapid stellar rotation, which would identify Apep as a rare candidate Galactic long-duration gamma-ray burst progenitor \citep{Callingham2019}. 

Here we report observations of the Wolf-Rayet binary, Apep, with the \textit{James Webb Space Telescope} (JWST) in the mid-infrared, targeting thermal emission from its warm ($\sim$100--300\,K) circumstellar dust. This paper provides an overview of the dataset and the dynamical and thermal record of the system contained within. The observations are described in Section~\ref{sec:observations} and the data reduction in Section~\ref{sec:reduction}. Section~\ref{sec:dynamical} analyses the structure and motion of the dust structure and Section~\ref{sec:thermal} models its thermal properties. Section~\ref{sec:discussion} discusses these new observational constraints and their implications for estimates of the distance to Apep. The findings are summarised in Section~\ref{sec:conclusions}. An accompanying study \citep{White2025} will present a model for colliding-wind dust structures applied to Apep to constrain its orbit and the role of a tertiary massive companion in further shaping the circumstellar dust structure. 

\section{Observations}
\label{sec:observations}

\begin{table*}[t]
    \centering
    \caption{Summary of JWST and VISIR data used in this study.}
    \label{tab:observing_log}
    \begin{tabular}{llllr}
    \hline
    Instrument & Filter & $\lambda_0$ / $\Delta\lambda$ ($\mu$m) & Calibrator & Date\\
    \hline
     JWST/MIRI & F770W & 7.7 / 1.95 & & 24 Jul 2024\\
     & F1500W & 15.0 / 2.92 & & 24 Jul 2024\\
     & F2550W & 25.5 / 3.67 & & 24 Jul 2024\\
     VLT/VISIR & J8.9 & 8.72 / 0.73 & HD\,178345 & 13 Aug 2016\\
     & & & HD\,133550 & 21 May 2018\\
     & & & HD\,133550 & 16 Jun 2024\\\
     & B11.7 & 11.52 / 0.85 & HD\,178345 & 23 Jul 2016\\
     & & & HD\,119193 & 18 Jun 2024\\
     & Q3 & 19.50 / 0.40 & HD\,145897 & 5 Jun 2018 \\
    \hline
    \end{tabular}
\end{table*}

\subsection{JWST/MIRI observations}
We observed Apep with the MIRI Imager onboard JWST on 24 Jul 2024 UT (programme ID 5842, \citealp{Han2024JWST}). The observations were carried out with full aperture direct imaging using broadband filters F770W, F1500W and F2550W. The FULL detector subarray was used for all observations to maximise the field of view while the FASTR1 readout pattern was selected to minimise saturation given the choice of subarray. For each filter, one exposure was taken at each of 12 dither positions under the default CYCLING dithering pattern and LARGE pattern size, starting at set number 5 and proceeding for 3 sets. The number of groups per integration was 28, 24 and 9 for the F770W, F1500W and F2550W filters respectively, which was set by the maximum number of groups before saturation in the secondary dust shell based on its predicted surface brightness from prior modelling if such a shell were to exist \citep{Han2020}. The number of integrations per exposure for the three filters was 1, 3 and 5 respectively, which was set by the total integration time required for detection of a potential secondary shell with the F770W filter and any shells close to the edge of the field of view within the F1500W and F2550W filters. An observing log is available in Table~\ref{tab:observing_log}. 

\subsection{VLT/VISIR observations}
We imaged Apep with the VISIR camera on the VLT's UT2 telescope at Paranal Observatory on two nights in June 2024. 
All observations were performed with full aperture. For each filter, 12 nodding cycles were performed to achieve an integration time of 2400s. Chopping was performed with an amplitude of 20$^{\prime\prime}$ and a position angle of 0, perpendicular to the nodding direction. Accompanying calibrator star observations were taken for flux calibration. An observing log is available in Table~\ref{tab:observing_log}.

\section{Data reduction}
\label{sec:reduction}

\subsection{JWST/MIRI data reduction}
The MIRI imaging data were calibrated with the JWST pipeline version 1.17.1 using the corresponding JWST Calibration Reference Data System context \texttt{jwst\_1322.pmap}. The Stage 3 products of each filter were used for structural and photometric analysis in this study. 

\subsubsection{Saturation recovery}
In all 3 filters, bright regions within the inner dust shell become saturated starting from group 2, preventing ramp fits for reliable flux estimates. However, inspection of individual frames along the integration ramp suggests that although the brightest pixels are saturated even among these group 1 frames (which prevents reliable photometry in these regions), the group 1 images do reveal significantly more structural detail of the inner shell compared to later groups in which larger areas of the inner shell become saturated. In an attempt to recover the structural (but not photometric) information of the inner shell from the group 1 images, we ran the Stage 1 pipeline by setting \texttt{n\_pix\_grow\_sat} to \texttt{0} in the \texttt{saturation} step and \texttt{suppress\_one\_group} to \texttt{False} in the \texttt{ramp\_fit} step. Default configurations were applied for all other pipeline steps to produce the Stage 3 data products. The data products from the pipeline with group 1 images obtained in such a way show noticeable artefacts which appear to be due to inaccurate flux extrapolations from the group 1 images nearing its saturated regions. To achieve improved smoothness for display purposes, we proceeded with custom processing as discussed below. 

For the only exposure at each dither location, the median of all group 1 readings among all integrations was taken for each pixel to constitute a group 1 image for the given exposure. Using the corresponding steps in the JWST pipeline, all such group 1 images were then matched with the World Coordinate System (\texttt{jwst.assign\_wcs.AssignWcsStep}) for resampling (\texttt{jwst.resample.ResampleStep}) to apply geometric corrections against optical artefacts. The resampled group 1 images at each dither location were aligned by fitting 2D Gaussian models to the bright background/foreground star (40$^{\prime\prime}$ from Apep at a position angle of 118$^\circ$) and the median image was taken. This median image was then calibrated into physical flux units by linearly scaling the image, such that the background level and the readings at the bright regions surrounding the saturated core align between the median group 1 image and the Stage 3 image for each of the three filters. The group 1 image was merged onto the Stage 3 image by applying a linear superposition between the two images, with the weight assigned to the group 1 image tapering from 1, for saturated pixels in the Stage 3 image, to 0 over the first-order to sixth-order neighbouring pixels. Note that while this procedure emphasises the structural detail and dynamic range achieved by these observations, the flux measurements within the inner dust shell are as not as reliable as those obtained by ramp fitting in the outer shells. We therefore only used this custom reduction for display purposes, not for quantitative analyses of the structure and emission of the dust shells. 

\subsubsection{Colour image}
\begin{figure*}
    \centering
    \includegraphics[width=\linewidth]{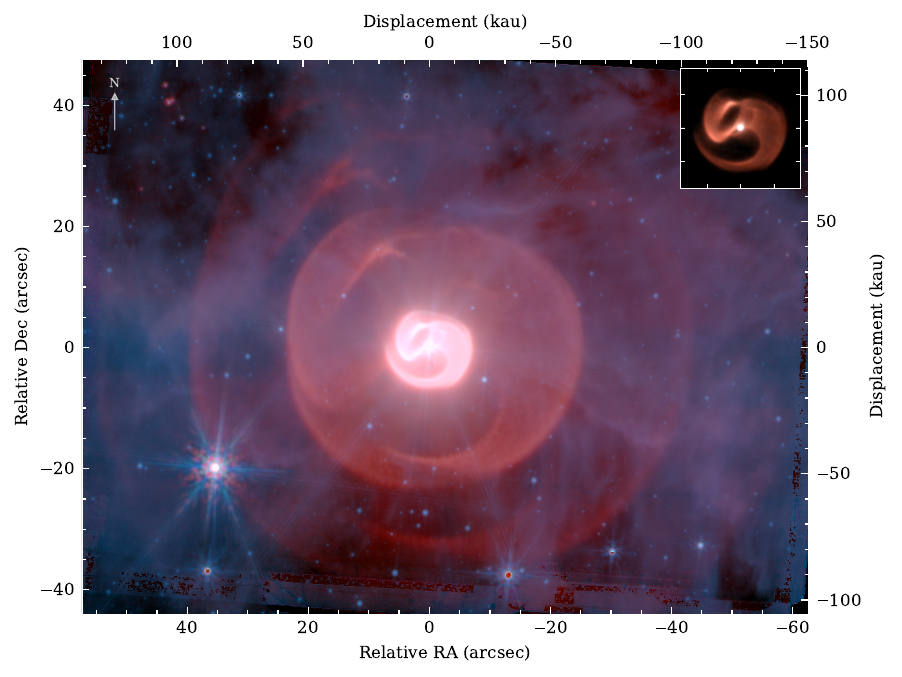}
    \caption{Colour image of Apep created from JWST/MIRI observations with the F770W, F1500W and F2550W filters. North is oriented towards the top of the page. A colour image of the inner dust shell observed with VLT/VISIR in the 2024 epoch is shown in the inset, where tick marks are separated by 5 arcsec. A distance of 2.4\,kpc is assumed for the linear displacement axes. }
    \label{fig:colour}
\end{figure*}

We created a false-colour image shown in Fig.~\ref{fig:colour} by combining the custom-processed images from the 3 JWST filters within the dataset. The images were first logarithmically transformed (with base 10 and an additive offset of 0.1~Jy/sr) before the background emission was subtracted (at levels of 1.0, 0.7 and 1.6 for the F770W, F1500W and F2550W filters in log space with units of Jy/sr) and the image floored to 0. Given the large dynamic range persistent in the log-transformed image, a further transformation was applied, with the image normalised to 1, by linearly mapping emission at levels defined by a series of reference points between 0 and 1 to an equal number of evenly spaced values within the same interval. These reference points were 0, 0.05, 0.10, 0.2, 0.4, 0.6, 0.8, 0.9, 0.95 and 1.0 for the F770W filter, 0, 0.05, 0.1, 0.2, 0.4, 0.6, 0.8 and 1.0 for the F1500W filter and 0, 0.05, 0.1, 0.2, 0.4, 0.6, 0.8, 0.9 and 1.0 for the F2550W filter. Using an RGB colour model, we assigned to the R channel 1.4 times the transformed F2550W image, the G channel 0.3 times the F1500W image plus 0.8 times the F770W image and the B channel 1.4 times the F770W image to obtain the final colour image. 

\subsection{VLT/VISIR data reduction}
The VISIR data were processed by first averaging image frames at each chopping location within a nodding cycle. Within each nodding cycle, two nodding frames were each obtained by subtracting the averaged image at one chopper position from the other within the given nodding position, and the chopped-nodded image was subsequently computed by subtracting the nodding frame at one nodding position from the other, resulting in four sub-images (two positive and two negative) within the frame. All sub-images in the chopped-nodded frame of all nodding cycles were aligned at the brightest pixel and averaged to produce a final image for each of the two filters. 

We calibrated detector readings to flux densities (in Jy/pixel) using the accompanying observations of calibrator stars. The database of flux densities of the mid-infrared standard stars used in the calibration is maintained by ESO\footnote{\href{www.eso.org/sci/facilities/paranal/instruments/visir/tools.html}{https://www.eso.org/sci/facilities/paranal/instruments/visir/tools.html}} based on \citet{Cohen1999}. 

To perform a proper motion analysis in this study, we also used archival VISIR observations from 2016 \cite{Callingham2019} and 2018 \citep{Han2020}. These observations are described in the corresponding references. We re-reduced these archival datasets in the same way as the 2024 epoch to ensure consistency in subsequent analyses.

\subsubsection{Colour image}
We created a false-colour image of the inner dust shell shown in Fig.~\ref{fig:colour} based on the two VISIR filters within the 2024 VISIR dataset. The images were first logarithmically transformed (with base 10 and an additive offset bringing the minimum surface brightness density in the image to 2.1~Jy/sr), before the median across the image was subtracted (in log space) and the image floored to 0 and normalised to 1. Using an RGB colour model, we assigned to the R channel 2 times the transformed B11.7 image, the G channel 1.4 times the J8.9 image plus 0.4 times the B11.7 image and the B channel 1.8 times the J8.9 image to obtain the final colour image.

\section{Dynamical evolution}
\label{sec:dynamical}

Our observations of Apep detect a series of 4 spiral dust shells centred on the binary, which extend to the edge of the MIRI field of view and correspond to a separation of $1.5 \times 10^5$\,au from the binary assuming a distance of 2.4\,kpc \citep{Callingham2019, Callingham2020, Han2020}. 
Owing to the varying wind-wind shock conditions resulting from modulation by the eccentric binary orbit \citep{Han2020, White2025}, dust production in Apep is episodic, creating cutoff regions and segmented edges inflated to large spatial scales by the expansion of dust. The resulting rich set of well-resolved substructures within the larger nested series of shells allows us to track in detail the dust evolution between successive formation events, where all except the innermost one were previously inaccessible from ground-based infrared imaging. We describe constraints on the dynamical evolution of dust derived from such observations in this section. 

\subsection{Measuring dust displacement}
\begin{figure}
    \centering
    \includegraphics[width=\linewidth]{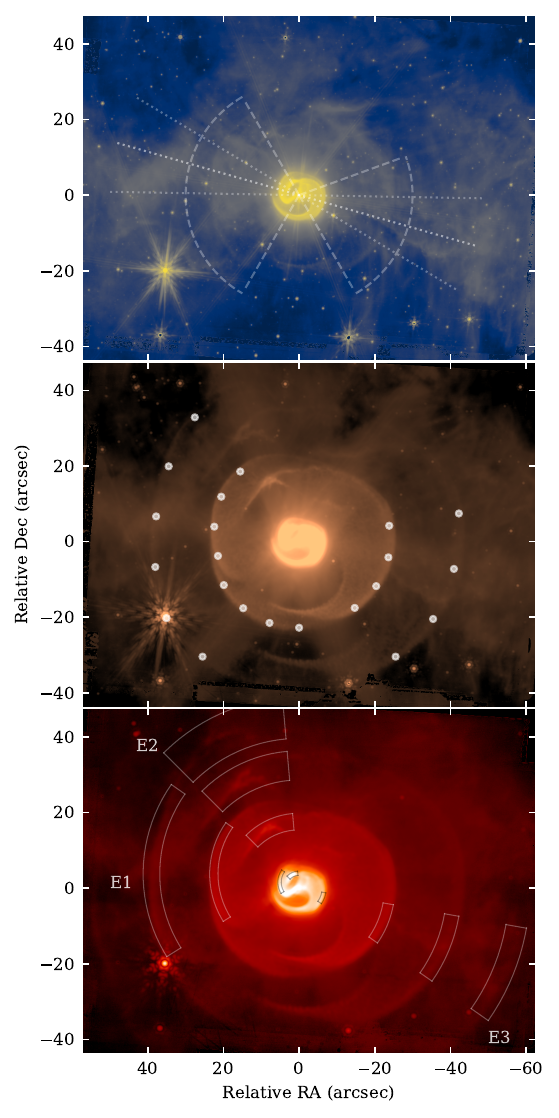}
    \caption{Top: MIRI F770W image of Apep with the azimuthal range of shell radius measured in this study indicated by dashed lines, and the direction orthogonal to the line of nodes and its uncertainty range \citep{White2025} indicated by dotted lines. Middle: F1500W image with points indicating examples of shell location detections. Bottom: F2550W image labelled with the E1, E2 and E3 features and apertures used to extract photometry.}
    \label{fig:filters}
\end{figure}

We measured the projected radial displacement of dust shells from the central binary in the 2 VISIR filters across 3 epochs and the 3 JWST filters in 1 epoch within a series of azimuthal bins (i.e., angular sectors with the apex at the binary) that partition one full revolution of the dust shell. 
The angular bins were each 20$^\circ$ wide, and measurements were only made in bins where the outer edge of a given dust shell reflects its deprojected separation from the binary. 
This is expected to be the case for the majority of outermost edges within a given episode of dust \citep{Han2020, Lau2020b}, as they arise from limb-brightening at the projected edges, except for the region to the north with dust absent from the otherwise azimuthally continuous distribution, possibly due to sculpting or destruction by a companion of the WR binary \citep{White2025}. 
We also avoided making measurements over a small azimuthal range to the south of the binary where the outer edge in this region is too dim for an accurate measurement in the inner shell observed with VISIR, and the southern edge of the third dust shell and beyond are outside the field of view of MIRI, preventing subsequent measurements of the expansion rate and shell spacing in this direction. The azimuthal range over which edge locations were measured are shown in Fig.~\ref{fig:filters}. 

To determine the location of edges within each azimuthal bin, we first emphasised edge-like features within each image by applying a high-pass filter. This was performed by subtracting a Gaussian-filtered image (with a standard deviation of 20 pixels in the JWST images and 3 pixels in the VISIR images) from the original image. We then derived the line profile of the high-pass-filtered image within each wedge. In practice, this was achieved by resampling the image in polar coordinates and averaging over each bin's azimuthal range. The edge location was then determined using the peak of the line profile. Examples of edge locations detected with this procedure are shown in Fig.~\ref{fig:filters}. 
As the innermost shell unavoidably saturates the MIRI detector, we carried out ground-based imaging with the VLT's mid-infrared camera, VISIR, separated by one month from the JWST observations, which we used to measure the inner shell's location.

\subsection{Shell spacing}
Given the high structural consistency across the VISIR and JWST filters, we used the inner shell's edge location determined from the 2024 VISIR epoch across both filters, the second shell's edge location determined from all 3 JWST filters and the third shell's edge location determined from the F1500W and F2550W filters to measure the spacing between neighbouring dust shells in each direction. Across all 9 azimuthal bins measured, we fitted the edge location as a linear function of the shell location with a Markov chain Monte Carlo (MCMC) approach implemented with the \texttt{emcee} package \citep{emcee}, leaving the uncertainties on the edge location as a free parameter. The fits for each azimuth are displayed in Fig.~\ref{fig:spacing_fits}. 

Across all azimuths with reliable measurements, we find that the radial dust displacement and shell order are closely described by a linear relation (see Fig.~\ref{fig:spacing_fits}), suggesting that dust expansion occurs with a highly stable velocity, unimpeded by potential sources of influence such as the interstellar medium even out to 0.6~pc from the binary (assuming a distance of 2.4\,kpc). Despite the differences in flux distribution between filters, the structural details are highly consistent across wavelength, which is expected from the optically thin emission and confined geometry of the thin dust surface. 

\begin{figure*}
    \centering
    \includegraphics[width=0.49\linewidth]{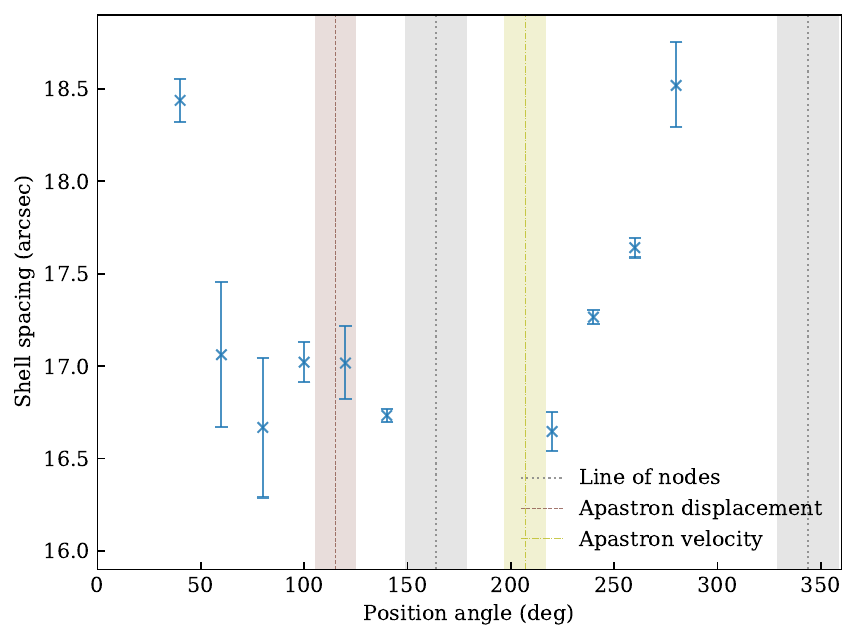}
    \includegraphics[width=0.49\linewidth]{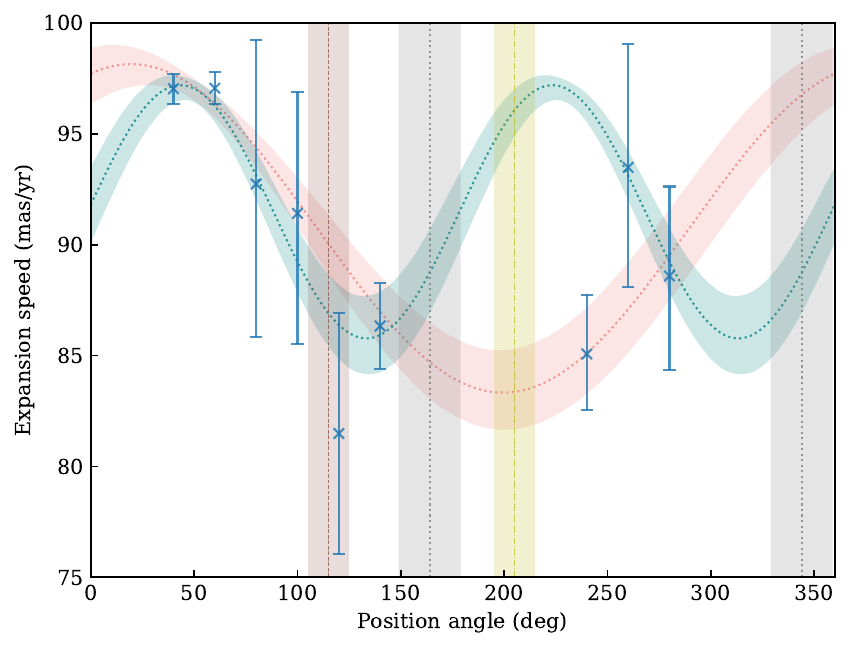}
    \caption{Left: Apep's inter-shell spacing as a function of position angle. The position angle in the horizontal axis corresponds to the angle measured counterclockwise from north in Fig.~\ref{fig:filters} (left panel), in which the azimuthal range with data points are labelled with arcs. Right: the expansion speed of Apep's inner dust shell as a function of position angle. The two curves correspond to the best-fit sinusoidal model with one and two periods over the full azimuthal range respectively. The vertical lines and shaded regions correspond to the line of nodes, apastron and the velocity vector at apastron and their uncertainties \citep{White2025}. }
    \label{fig:shells}
\end{figure*}

Despite the smoothness of the structures, the outflow is not circularly symmetric. Fig.~\ref{fig:shells} displays the inter-shell spacing that we derived as a function of azimuth, which shows deviations from uniformity modulated by 11\% at its peak relative to trough within the azimuthal ranged accessible. 
To further ascertain the non-uniform expansion of dust, we collected archival VLT/VISIR observations of the inner dust shell between 2016 and 2018 to measure the dust expansion rate as described in the following section. 

\subsection{Proper motion}
The inner dust shell has previously been observed by VISIR on 3 occasions between 2016 and 2018. The addition of the 2024 epoch quadruples the time baseline over which to track the expansion of the dust, which we leverage to yield significantly more reliable proper motion constraints than previous work that relied on sub-pixel edge detection methods \citep{Han2020}. Note that we did not include the 2017 epoch in this analysis due to the chopping and nodding configurations used in that epoch which were not optimised for reliably imaging the extended structures in Apep. 

\begin{figure}
    \centering
    \includegraphics[width=0.9\linewidth]{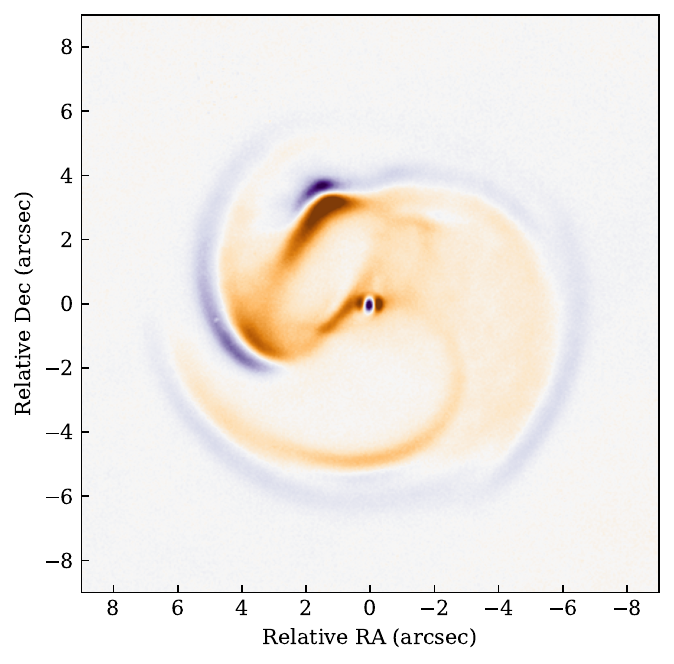}
    \caption{The difference image of the inner shell of Apep obseverved by VISIR with the B11.7 filter between 2016 and 2024. Over-emission in 2024 compared to 2016 is shown in purple and under-emission is shown in orange. The difference image suggests a consistent and significant expansion of the dust structure which we measured to be $90 \pm 4$\,mas\,yr$^{-1}$. }
    \label{fig:subtraction}
\end{figure}

Comparison between the 2016 the 2024 VISIR epochs shows clear measurable expansion of the inner shell, as shown in Fig.~\ref{fig:subtraction}. We used the edge locations determined from the 2016, 2018 and 2024 VISIR epochs in both the J8.9 and B11.7 filters where available to measure the expansion speed of the inner dust shell in each direction. Across all 8 azimuthal bins with reliable measurements, we used an MCMC to fit a linear function relating edge location to the date of observation, where the slope was used to determine the expansion speed. The astrometric uncertainty on the edge location was included as a free parameter. The radius as a function of time for each azimuth fitted is displayed in Fig.~\ref{fig:pm_fits}. The resulting expansion speeds measured are displayed as a function of azimuth in Fig.~\ref{fig:shells}. 

We find that the expansion speed exhibits a similar azimuthal modulation to the shell spacing. To estimate the amplitude of this variation, we fitted a sinusoidal model defined by
\begin{equation}
    v(\theta) = \frac{1}{2}(1 + \xi)v_\mathrm{min} + \frac{1}{2}(1 - \xi)v_\mathrm{min} \cos \left( \theta - \theta_\mathrm{min} \right ),
\end{equation}
where $v$ is the expansion speed, $\theta$ is the position angle, $v_\mathrm{min}$ is the minimum expansion speed, $\xi$ is the maximum expansion speed as a multiple of $v_\mathrm{min}$ and $\theta_\mathrm{min}$ is the position angle at which $v = v_\mathrm{min}$. Sampling the ($v_\mathrm{min}$, $\xi$, $\theta_\mathrm{min}$) parameter space with an MCMC, we found sinusoidal fits suggests a minimum expansion speed of $v_\mathrm{min} = 83 \pm 2$\,mas\,yr$^{-1}$ at $\theta_\mathrm{min} = 199 \pm 11$ and a $\xi - 1 = 18 \pm 3$\% faster expansion at maximum. The mean expansion speed across all azimuths measured is $90 \pm 4$\,mas\,yr$^{-1}$.

\subsection{Orbital period}
\label{sec:period}
Although the shell spacing appears to vary across azimuth, it is found to be highly consistent between neighbouring dust shells along a fixed azimuth, suggesting that the expansion speed is constant for each dust shell along a given azimuth. As one dust shell is thought to be produced from each orbital period of the WR binary, we used the inner shell expansion speed and the shell spacing measured for each azimuthal bin to determine the orbital period of the binary. The scatter between the periods measured from each azimuthal bin show no clear trend and lie within an interval between 175 and 210\,yr, and is likely a result of the larger fractional uncertainties on the expansion speed, which is determined from more closely spaced structures (the expansion of one shell, rather than the spacing between shells), compared to the shell spacing measurements. We therefore estimated the period of the binary using the mean across all azimuths measured and its uncertainties the scatter between azimuths, giving a value of $193\pm11$\,yr. This implies that the dust structure detected within the MIRI field of view reflects approximately 700~yr of dust production history and contains the most ancient dust structures resolved around WR binaries, serving as a fossil record of the dynamical and thermal conditions in the system inflated by its steady expansion.

\section{Thermal evolution}
\label{sec:thermal}
 
The combined JWST and VLT datasets offer exceptional dynamic range and wavelength coverage to track the thermal evolution of dust out to at least $10^5$\,au, which has previously been inaccessible from ground-based infrared observations. In this section, we use the photometric information in the multi-wavelength images to model the temperature of dust as a function of radius.

\subsection{Photometry}
As the several structural features within each dust shell expand at different rates in projection, their emission overlaps in the image at higher order shells, making it difficult to separate the flux contribution from each shell at an arbitrary location in the image. Instead of measuring the emission of entire dust shells, we selectively measured the flux density of features along azimuths that are minimally impacted by overlapping shells, the diffuse filamentary nebula, the PSF of the bright core and background/foreground stars. 

We selected 3 features along widely separated azimuths for which bright and consistent emission could be photometrically characterised across a wide span of shell orders and wavelengths. These features include an eastern outer arc on the bright ellipse (E1), a northern cusp on an inner edge (E2) and a western outer arc on the tail of the dust shell (E3), which are indicated in Fig.~\ref{fig:filters}. We extracted the flux density of E1, E2 and E3 in shells 2 to 4 (with shell 1 being the inner shell) from all 3 JWST filters, and in shell 1 from all 3 VISIR filters across 6 images taken across 2016, 2018 and 2024. For each feature and image combination, background emission was determined using an annulus outside the outmost shell in which the feature was detected and within the same azimuthal range as the corresponding feature, and was subtracted from the flux densities measured for the given feature. Extinction correction correction was applied assuming $A_v = 11.4$ \citep{Callingham2019, Han2020} and the extinction law determined by \citet{Fitzpatrick2007}. Although these apertures have been selected to minimise flux contamination, the photometric uncertainties for these specific features are still dominated by flux contributions from the faint background or foreground dust structures and the extended nebula, which we estimate to be 10\%. Our calibrated flux densities for E1, E2 and E3 across the dataset are shown in Fig.~\ref{fig:sed_fits}.

\subsection{Resolved SED modelling}
\label{sec:sed}
We modelled the SED of the E1, E2 and E3 features assuming that dust expands radially over time and consistently from inner to outer shells, such that the region of emission captured within the aperture can be interpreted as temporal evolution of the same matter. 
This enabled us to fit a dust emission model to the spatially resolved spectral energy distribution (SED) over the two-dimensional grid of E1 to E3 features and their appearance captured at different radii, as shown in Fig.~\ref{fig:sed_fits}. Although no individual SED is captured by all 5 filters, the fact that they trace the same feature at different points in space and time significantly restricts the range of possible dust temperatures at each location to build a self-consistent model. 

For each feature, we assumed optically thin emission described by
\begin{equation}
\label{eq:fnu}
F_\nu = Q_\mathrm{abs}(a, \lambda) B_\nu(\lambda, T) \Omega_\mathrm{eff},
\end{equation}
where $F_\nu$ is the flux density, $Q_\mathrm{abs}$ is the absorption coefficient, $a$ is the dust grain diameter, $\lambda$ is the wavelength, $B_\nu$ is the Planck function, $T$ is the temperature and $\Omega_\mathrm{eff} = \int \tau d\Omega$ is the effective angular area of emission, where $\tau$ is the optical depth and $\Omega_\mathrm{eff}$ is the angular area within the aperture covered by the emitting structure. Wolf-Rayet binary dust has been found to be $\lesssim$0.1\,$\mu$m in size \citep{Lau2023}, such that $a \ll \lambda$ in the mid-infrared. We therefore estimated $Q_\mathrm{abs}$ assuming
\begin{equation}
    \label{eq:qabs}
    Q_\mathrm{abs} = Q_0 \left( \frac{\lambda_0}{\lambda} \right)^{\beta},
\end{equation}
setting $\beta = 1$ for amorphous carbon dust \citep{Rouleau1991, Cherchneff2000}, and normalising the function with $Q_0 = 1$ and $\lambda_0 = 1$\,$\mu$m. Note that the specific choice of normalisation does not affect constraints on $T$, as any deviations can be absorbed by its degeneracy with the nuisance parameter, $\Omega_\mathrm{eff}$, during model fitting. 

\begin{figure*}
    \centering
    \includegraphics[width=0.7\linewidth]{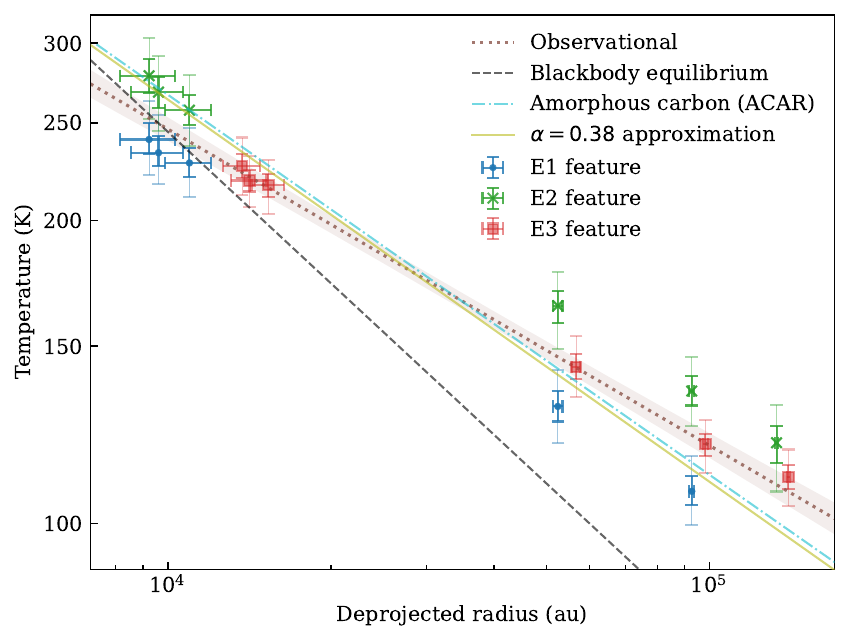}
    \caption{The radial temperature profile of dust in Apep measured from its spatially resolved SED. The thicker error bars are derived by fitting SED models to individual features. The thinner error bars show the adjusted uncertainties upon enforcing the condition that the measurements sample the same temperature profile. The best-fit power law profile with a power-law index of -0.31 is shown, along with the blackbody equilibrium temperature, an amorphous carbon equilibrium temperature profile based on \texttt{RADMC-3D} simulations and an amorphous carbon approximation with a power law index of -0.38 for comparison.}
    \label{fig:temperature_profile}
\end{figure*}

Given the consistent structures seen across wavelength and shells, we assumed that $\Omega_\mathrm{eff}$ remains constant for a given feature across all dust shells and the time of observation, as appropriate in the optically thin regime (i.e., the optical depth decreases, but the integrated effective angular area of emission remains constant as the dust structure expands). For each dust feature, we assigned as free parameter the temperature of the feature found at each radial location. These radial locations include the inner dust shell at its 2016, 2018 and 2024 locations observed by VISIR, and shells 2 to 4 at their 2024 locations observed by JWST, resulting in the each feature sampled at a maximum of 6 radial locations. This results in the parameter space of ($\Omega_\mathrm{eff}$, $T_1$, $T_2$, ..., $T_n$) to be sampled for $n$ SEDs of each feature, where $n = 6$ for E2 and E3 and $n = 5$ for E1 (since E1 is barely detected in the fourth shell of any image in the dataset). We sampled this parameter space for each feature with an MCMC, the results of which are displayed as a function of radius in Fig.~\ref{fig:temperature_profile}, with the individual SED fits shown in Fig.~\ref{fig:sed_fits}. Note that since $\Omega_\mathrm{eff}$ is fixed for each feature, temperature constraints can be obtained at a given radius even with photometry at only one wavelength if the feature becomes faint at this radius at all other wavelengths, which relies on $\Omega_\mathrm{eff}$ constraints provided by the same feature sampled at other radii. 

With dust location measurements made for each outer shell and for the inner shell at each epoch, the individual resolved SEDs form a temperature ladder, enabling us to trace the cooling curve of the dust as it expands, as shown in Fig.~\ref{fig:temperature_profile}. E1 and E3 are limb-brightened outer edges in the sky plane, hence their projected radius from the binary measured at the appropriate position angle is the same as their deprojected radius. E2 is seen in projection, and belongs to the bright ellipse to which E1 is attached, thus they share approximately the same deprojected radius. The three features together further sample dust at different radial locations more densely than any feature alone does. 

\subsection{Temperature profile}
\label{sec:temperature_profile}
The temperature profile shown in Fig.~\ref{fig:temperature_profile} closely follows a power law. We fitted such a parametrisation to the temperature profile, defined by
\begin{equation}
\label{eq:tpl}
    T(r) = T_0 \left( \frac{r}{r_0} \right)^{-\alpha},
\end{equation}
where $r_0$ was set to 10,000\,au (such that $T_0$ is the temperature at 10,000\,au). Although the temperature profiles found for the 3 features show differences just beyond uncertainties, given potential biases in photometric calibration, we do not at this stage interpret these differences to be due to real physical mechanisms. Instead, they likely reflect underestimated uncertainties returned by fitting the SEDs of individual dust features. For example, E1, E2 and E3 are each affected to a different extent by outer dust structures that overlap in projection, which could lower the apparent temperature, and the azimuthally asymmetric diffraction features of the bright core, which could make the temperature appear hotter. Given the reasonably well-understood geometry of the system \citep{White2025}, uncertainties on the temperature likely dominate over uncertainties on the radius. Under the assumption that the cooling law is isotropic and that the temperature at similar deprojected radii should be approximately the same, we assumed that the temperature uncertainties fitted from individual SEDs are underestimated by a factor of $\zeta$, and we included $\zeta$ as a free parameter when fitting the temperature profile. We fitted the power law temperature profile with an MCMC to all sample points regardless of the feature to which they correspond, finding that $\alpha = 0.31 \pm 0.02$ and $T_0 = 246 \pm 7$\,K. The fitted temperature profile and the adjusted uncertainties are shown with the thinner error bars in Fig.~\ref{fig:temperature_profile}. We also plotted SED models for each feature corresponding to this best-fit power-law temperature profile in Fig.~\ref{fig:sed_fits}, which show reasonable fits to the SEDs and support the assumption that the 3 features follow the same temperature profile.

The power law index that we derived suggests that the radial temperature profile of dust around Apep falls off significantly more slowly than the blackbody equilibrium temperature. This likely reflects significantly different absorption properties of WR dust compared to blackbody. 
To compare the slope with the expected value under the approximation in Eq.~\eqref{eq:qabs}, we assumed that the grains are in thermal equilibrium without scattering, such that the emission and absorption coefficients are equal. Grains in thermal equilibrium then satisfy
\begin{equation}
    \label{eq:eq}
    \int Q_\mathrm{abs}(a, \lambda) B_\nu(\lambda, T) d\nu = \langle Q_\mathrm{abs} \rangle_* L_* \frac{\pi a^2}{16 \pi r^2},
\end{equation}
where $\langle Q_\mathrm{abs} \rangle_*$ is the mean absorption coefficient weighted by the stellar spectra. We further assumed that the main contributor to the stellar luminosity is emission at short wavelengths compared to the grain size, where the absorption efficiency can be approximated as 1, such that $\langle Q_\mathrm{abs} \rangle_*$ can be set to 1. Under the $1/\lambda$ approximation for $Q_\mathrm{abs}$ in Eq.~\eqref{eq:qabs} appropriate for the wavelength range over which the grain emission predominantly lies, the solution to Eq.~\eqref{eq:eq} is given by
\begin{equation}
    \label{eq:mbb}
    T = \left( \frac{a^2 h^4 c^3}{32 \pi AQ_0 \lambda_0 k^5} \right)^{1/5} \frac{L_*^{1/5}}{r^{2/5}},
\end{equation}
where $A = \int_0^\infty x^4 / (e^x - 1) dx \approx 24.9$. This theoretical temperature profile that scales with $r^{-0.4}$ thus deviates from the $L_*^{1/4}$ and $r^{-1/2}$ dependence for blackbody grains.

A profile shallower than the blackbody equilibrium temperature has also been assumed in prior work in the context of WR140 \citep{Williams2009}, where a power law with $\alpha = 0.38$ has been assumed for amorphous carbon dust grains. While such a slope is still shallower than the value that we observationally derived, it offers a significantly improved fit than the blackbody temperature profile, suggesting that the thermal properties of dust grains in the system could be reasonably described by amorphous carbon grains within the precision achieved with these observations.

\section{Discussion}
\label{sec:discussion}

\subsection{Outflow asymmetry}
In this section, we discuss some possible scenarios that could be proposed to explain the azimuthal asymmetry in dust expansion observed. 

\subsubsection{Wind modulation by an eccentric orbit}
One possible explanation for an azimuthally asymmetric dust expansion is that the post-shock outflow velocity could vary across the binary's orbit, which has been inferred to be eccentric \citep{Han2020, White2025}. In WR140, extreme orbital modulation has been found to almost completely suppresses dust production near periastron \citep{Han2022}. It is conceivable that even within an episode of dust formation, modulations could be manifest not only in the dust density, but also in its geometry. 
Under the orbital modulation scenario, the extrema in the outflow velocity likely align with periastron/apastron, the projected position angle of which is labelled in Fig.~\ref{fig:shells} based on the orbit derived in \citet{White2025}, which also aligns approximately with the median position angle of the dust structure as expected \citep{Williams2019}. This comparison shows that smaller expansion speed and shell spacing align more closely with apastron. 

Variations in the wind momentum ratio could affect the opening angle of the bow shock \citep{Gayley2009}, however the opening angle alone cannot explain variations in dust outflow speed as it does not affect the radial displacement of the dust's outer edge. Faster expansion of dust launched at periastron is also contrary to the expected effect if both stellar winds were subject to the same fractional suppression in outflow speed due to radiative braking, which would reduce stellar wind speeds at periastron relative to the rest of the orbit. Hydrodynamical modelling of the wind shock is required to test whether post-shock velocities could indeed be higher at smaller binary separations.

\subsubsection{Velocity contribution from orbital motion}
It is possible that gravitational reflex motion near periastron could increase the outflow speed.
Assuming an orbital eccentricity of 0.8 \citep{White2025}, an orbital period of 193\,yr derived in Section~\ref{sec:period}, the masses of the binary components are equal and a semi-major axis of 67\,au (based on the binary angular separation and position angle, \citealp{Han2020}, and a distance of 2.4\,kpc), the orbital speed at periastron in the centre-of-mass frame is approximately 40\,km\,s$^{-1}$ or 4.5\,mas\,yr$^{-1}$ at a distance of 2.4\,kpc. For an unequal-mass binary, this value could increase by up to a factor of 2, which would just be comparable to the amplitude of modulation that we observe. Under this scenario, we expect the position angle of the fastest dust outflow to align with the projected velocity vector at periastron. 
We indeed observe that the minimum outflow speed aligns closely with the projected position angle of apastron, as shown in Fig.~\ref{fig:shells}. It would therefore appear feasible for orbital motion to contribute to the asymmetric outflow observed under these estimates.

\subsubsection{Interactions with the interstellar medium}
\label{sec:asymmetric_winds}
Another scenario that might be proposed is asymmetries introduced by interactions with the interstellar medium. The interstellar medium would be required to be dense ($\sim10^4 \, \mathrm{cm}^3$) in such a scenario to reshape the dust motion. Under the Epstein gas drag regime, which applies to diffuse gas like in the interstellar medium and small ($\sim0.1\,\mu$m, \citealp{Lau2023}) dust grains as seen in WR stars, the gas stopping time for an interstellar medium density of 100 cm$^{-3}$ is 20\,kyr, which is two orders of magnitude larger than the age of the dust observed, even when assuming the gas is heated to 150\,K by the WR binary. In order for the gas-drag timescale to be comparable to the age of the dust, the system must be situated and moving relative to a two orders of magnitude denser region of gas than estimated above. This is disfavoured given that even if the dust grains were to survive sputtering as they move through such a dense ISM, such interactions should result in further deceleration of the dust during its expansion, which we do not observe.

\subsubsection{Non-spherical wind velocity field}
Although the lack of edges expanding in the sky plane in certain azimuths prevents a full characterisation in all directions, the shell spacing measurements could accommodate the possibility of two periods of oscillations per revolution, especially considering that such variations in azimuth may not be sinusoidal or symmetric. One scenario that could manifest as such a trend is if the dust expansion is spherically asymmetric, specifically with a velocity field that is prolate, which would appear circularly asymmetric when viewed at a nonzero inclination relative to a pole-on perspective. It has previously been suggested, based on measurements of the sky-plane dust expansion speed \citep{Callingham2019, Han2020} and the line-of-sight spectroscopic wind speed \citep{Callingham2019, Callingham2020}, that the wind launched by a WR component in Apep could be anisotropic by a factor of 4 \citep{Callingham2019, Han2020}. The exact azimuthal variation of the dust expansion seen in projection depends on the latitudinal distribution of the velocity field, but if the projected radial expansion varies sinusoidally, then a 24$^\circ$ inclination in the case of Apep \citep{White2025} would result in an approximately 12\% faster peak projected expansion than at minimum for an outflow that is 4 times faster in the polar direction relative to the equator, assuming that the axis of symmetry of the outflow aligns with the orbital axis. This is consistent with the level of asymmetry observed in Apep. The direction of minimum expansion speed also aligns approximately along the line of nodes (see Fig.~\ref{fig:filters}), which corresponds to the two directions from the binary that would reflect a slow equatorial outflow rate under such a scenario. 

Although dust could form at a velocity slower than the terminal wind speed in the colliding-wind region, radiation pressure and/or secondary wind interaction downstream from the initial shock is expected to subsequently accelerate the dust (\citealp{Williams2009, Han2022}, Monnier et al., submitted) such that the post-shock dust eventually reaches a speed comparable to the pre-shock wind speed. Indeed, the dust expansion speed observed in other colliding-wind Wolf-Rayet binaries largely agree with the wind speed \citep{Tuthill2008, Lau2020b, Williams2009, Han2022}. The non-spherical outflow scenario in Apep \citep{Callingham2019} was suggested to explain a factor of 4 discrepancy between the sky-plane expansion speed of dust \citep{Han2020} and the line-of-sight wind speed measured from spectroscopy \citep{Callingham2020}. It has been proposed that rapid rotation could support the large latitudinal wind speed gradient, implying a WR component in Apep is a Galactic long-duration gamma-ray burst progenitor. While previous estimates of the dust expansion rate have relied on sub-pixel edge detection methods to search for small shifts in dust structure and carry large uncertainties, the quadrupled time baseline enabled by the 2024 epoch over which to measure this expansion would confirm the dust expansion speed to be four times lower than the wind speed of the WN component if the distance to Apep is 2.4\,kpc. Note that the distance may be further than 2.4\,kpc, which would reduce this discrepancy, as we discuss in Section~\ref{sec:distance}. 

For context, the half-opening angle of the shock cone is determined to be $63 \pm 3^\circ$\citep{White2025}, and the wind momentum ratio $0.44 \pm 0.10$ \citep{Marcote2021} assuming that the shock is adiabatic. Assuming the orbit determined by \citet{White2025} and distance of 2.4\,kpc, the periastron separation between the binary is $12\pm2$\,au. Alternative explanations for the wind speed discrepancy have also been proposed, which include a deceleration of dust expansion following its formation in the fast wind. However, this scenario is firmly rejected by the dust structures visible in the MIRI images, which are highly regular in spacing and suggest outflow at a speed that is uniform over time. It has also been proposed that the wind speeds could be highly variable, and that the fast wind speed was measured at a time of accelerated outflow. The regular spacing of the shells and its well-preserved substructures from one episode of dust to the next also argues against wind speed variations on the timescale of hundreds of years, which would have resulted in irregularly spaced shells, with faster winds destroying the regularity of the intricate structures observed.  
Either a non-spherical outflow or a further distance is therefore the more plausible explanation in light of the MIRI images and the asymmetric expansion that we find.

\subsection{Constraints on luminosities}
\label{sec:luminosity}
While extinction from the extremely dusty environment of Apep has made it difficult to measure the luminosity of the binary, this dust also offers an avenue to such a measurement through its thermal emission. While the degeneracy between the absorption coefficient, $Q_\mathrm{abs}$, and integrated optical depth, $\Omega_\mathrm{eff}$, did not prevent temperature fitting (Eq.~\ref{eq:fnu}), inferring the stellar luminosity from the dust temperature profile requires the normalisation of the absorption coefficient to be realistically estimated. 

To constrain $Q_\mathrm{abs}$ and compare the temperature profile derived in this study with realistic dust grain properties in thermal equilibrium, we performed radiative transfer simulations with the \texttt{RADMC-3D} code \citep{Dullemond2012} based on laboratory-measured optical constants for amorphous carbon grains \citep{Zubko1996} assuming a grain size of 0.1$\mu$m. We used optical constants measured for ACAR amorphous carbon grains determined by \citep{Zubko1996} based on laboratory measurements and computed dust opacities with Mie theory\footnote{Code based on RADMC-3D \\ \href{gitlab.mpcdf.mpg.de/szucs/bh-mie-scat/-/tree/master}{https://gitlab.mpcdf.mpg.de/szucs/bh-mie-scat/-/tree/master}} assuming a grain size of 0.1~$\mu$m. We computed the radial dust temperature profile by placing 3 stellar components at the centre of the spherical grid with temperatures of 79,400\,K, 63,100\,K and 36,000\,K, corresponding to typical temperatures for the WN4-6b, WC8 and O8\,Iaf \citep{Callingham2020} components respectively. Note that while in reality the O8 Northern Companion is offset from the WR binary by 0.7$^{\prime\prime}$ in projection, the effect of the offset on the segment of temperature profile measurable is expected to be insignificant, since all 3 features, at different locations on the dust shells, follow approximately the same temperature profile and the scatter between them is interpreted as due to uncertainties. 

The stellar radii were adjusted such that the same luminosity ratio between the three stellar components was maintained and only their total luminosity was varied. This luminosity ratio was chosen to be consistent with the near-IR interferometric constraints and the near-IR SED assembled from VLT/NACO observations in \citet{Han2020}, which was corrected for interstellar extinction assuming $A_v = 11.4$ \citep{Callingham2019}. To calibrate flux density to luminosity, we used WR and O star spectra from the Potsdam WR spectral database \citep{Hamann2004, Sander2012, Hainich2019}, including a Galactic WC model at 63,100\,K with $R_t = 1 \, \mathrm{R}_\odot$, a Galactic WNE model at 79,400\,K with $R_t = 1 \, \mathrm{R}_\odot$ and a Galactic O star model at 36,000\,K with log($g$)[cgs] = 3.8. We used photometry from the J, H, Ks, IB\_2.24, NB\_3.74 and NB\_4.05 filters to constrain the flux of the northern companion, but only the J-band flux for the WR binary as its SED suggests the presence of close-in hot dust which contaminates the stellar flux at longer wavelengths \citep{Han2020}. We find that a WC:WN luminosity ratio of 2.7:1 is required to reproduce the near-IR binary contrast of 5:1 \citep{Han2020}, and a WR binary to O star luminosity ratio of 1.5:1 is required to reproduce their near-IR flux density ratios \citep{Han2020}. Note that in the near-IR, the SED could be contaminated by free-free emission, thereby biasing the luminosity estimates to be higher than the true value. However, the effect of any free-free contamination is likely limited given the broad consistency of the near-IR SED with the XSHOOTER spectrum at visible wavelengths \citep{Callingham2020} without incorporating a free-free emission component \citep{Han2020}. 

We find that to reasonably reproduce the temperature profile under a distance assumption of 2.4\,kpc, the combined luminosity of the binary is required to be $6.0^{+0.7}_{-0.5} \times 10^5\,\mathrm{L}_\odot$. The temperature profile simulated for such a luminosity is shown in Fig.~\ref{fig:temperature_profile}, which is closely described by the aforementioned $\alpha = 0.38$ power law approximation. A further distance assumption would require a larger luminosity, which we discuss in Section~\ref{sec:distance}.

\subsection{Filamentary nebula}
\label{sec:filament}
 
In addition to the regular concentric dust shells, the MIRI images also suggest the presence of a diffuse filamentary nebula, which is most clearly visible in the F770W filter (see Fig.~\ref{fig:filters}). The fact that the nebula becomes decreasingly visible towards longer wavelengths relative to the regularly shaped spiral dust shells suggests that the filamentary nebula exhibits either strong spectral features captured at shorter wavelengths or a significantly hotter temperature. Diffuse dust structures have been observed around other WR binaries \citep{Richardson2025}, however their spectral slopes appear to be redder that observed here, although quantitative analyses of those systems is required to confirm this comparison. We performed aperture photometry of the filamentary nebula within a region of significant emission, subtracting off background emission nearby that appear largely devoid of its presence. If the filamentary dust structure is seen in primarily thermal continuum emission, we find that reproducing its negative spectral slope between the three MIRI wavelengths would require a dust temperature of at least 350\,K at a projected separation of $10^5$\,au, which is significantly hotter than the spiral dust shell at a comparable radius even if it lies within the sky plane. The different spectral slopes between the filamentary nebula and the spiral dust shells cannot be explained by the former being significantly closer to us along the line of sight, as the interstellar dust extinction levels for Apep is only a few percent at these wavelengths. 

\begin{figure*}
    \centering
    \includegraphics[width=0.8\linewidth]{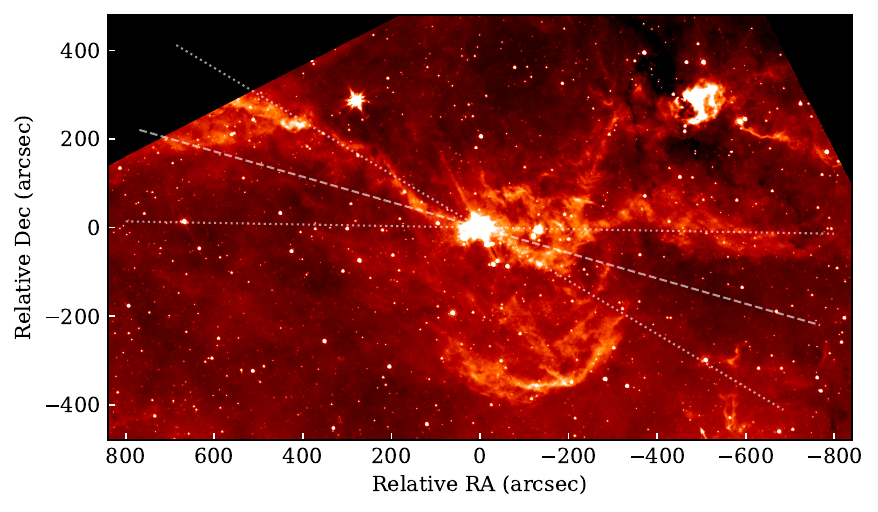}
    \caption{\textit{Spitzer} observations of Apep at 7.9\,$\mu$m. The direction orthogonal to the line of nodes \citep{White2025} is indicated with dashed lines and uncertainty range with dotted lines. }
    \label{fig:spitzer}
\end{figure*}

Even if the filamentary nebula emits strong PAH features captured by the F770W filter that make it significantly brighter at these shorter wavelengths, this would still leave open the question of whether the nebula exists as a foreground/background object as part of the interstellar medium, or whether it is associated or even at least partly produced or dynamically or thermally shaped by Apep. The orientation of the filament appears to broadly align with the direction orthogonal to the binary's line of nodes (see Fig.~\ref{fig:filters}), which could make it coincident with the polar axis of the WR stars if the filament extends significantly along the line of sight. \textit{Spitzer} observations at similar wavelengths have also detected the filament out to at least 10\,pc in projection (see Fig.~\ref{fig:spitzer}), which also shows broad alignment with Apep's projected orbital axis, in addition to further dust arcs to the south. Chance alignment of such a prominent structure traversing right through a rare double WR binary within this field of view appears to be unlikely, although for massive stars that evolve over short timescales, it is also plausible for ISM filaments from their formation to remain in their vicinity. 

Although this probability argument alone is not sufficient to establish association, we could speculate on possible mechanisms in the scenario that the filament is being shaped or influenced by the central binary that are worthy of further investigation. Evidence of helical outflows has been observed around massive, evolved stellar environments \citep{Lau2016}, and PAH features have been suggested in the outflow of WR stars \citep{Marchenko2017}. Furthermore, the potential rapid spin in Apep could conceivably generate magnetic fields via stellar dynamos, channelling ionised particles into polar outflows. Assuming a polar outflow four times as fast as the dust shell expansion rate and similar to the WN wind speed, and a polar axis inclination equal to that of the orbital axis at 24$^\circ$ \citep{White2025}, the dust filament in the Spitzer frame must have formed over a timescale of approximately 5,000\,yr. Under such a scenario, the helical shape of the filament could reflect past orbital precessions of the binary, possibly due to a companion star detected to the north of the binary \citep{Callingham2019} via the Kozai-Lidov mechanism \citep{White2025}. \citet{White2025} estimated the precession timescale to be on the order $10^5$\,yr, although significant uncertainties on the mass and orbit of the companion star could allow for faster precession timescales via the eccentric Kozai mechanism. Follow-up studies that investigate the composition and motion of the dust filament relative to Apep are required to establish the connection between the two. 

\subsection{Distance to Apep}
\label{sec:distance}

While an anisotropic wind velocity field has been suggested to explain the difference between the sky-plane dust expansion speed and line-of-sight wind speed (see Section~\ref{sec:asymmetric_winds}, \citealp{Callingham2019}), an alternative possibility is that the distance to Apep is significantly larger than previously estimated.
If Apep's distance is underestimated by a factor of 2 to 3, this would bring the system in line with other WR binaries for which the two measurements are approximately consistent \citep{Richardson2025}. The distance of Apep has been difficult to determine from Gaia astrometry due to its binary nature \citep{Callingham2019}. Apep has also not been identified as a member of an OB association \citep{Callingham2019}. Dropping the 2.4\,kpc assumption used so far in this study, the temperature profile, which constrains the stellar luminosity as a function of radius in au terms, offers constraints on the distance from a thermal perspective. 

\begin{figure*}
    \includegraphics[width=1.0\linewidth, right]{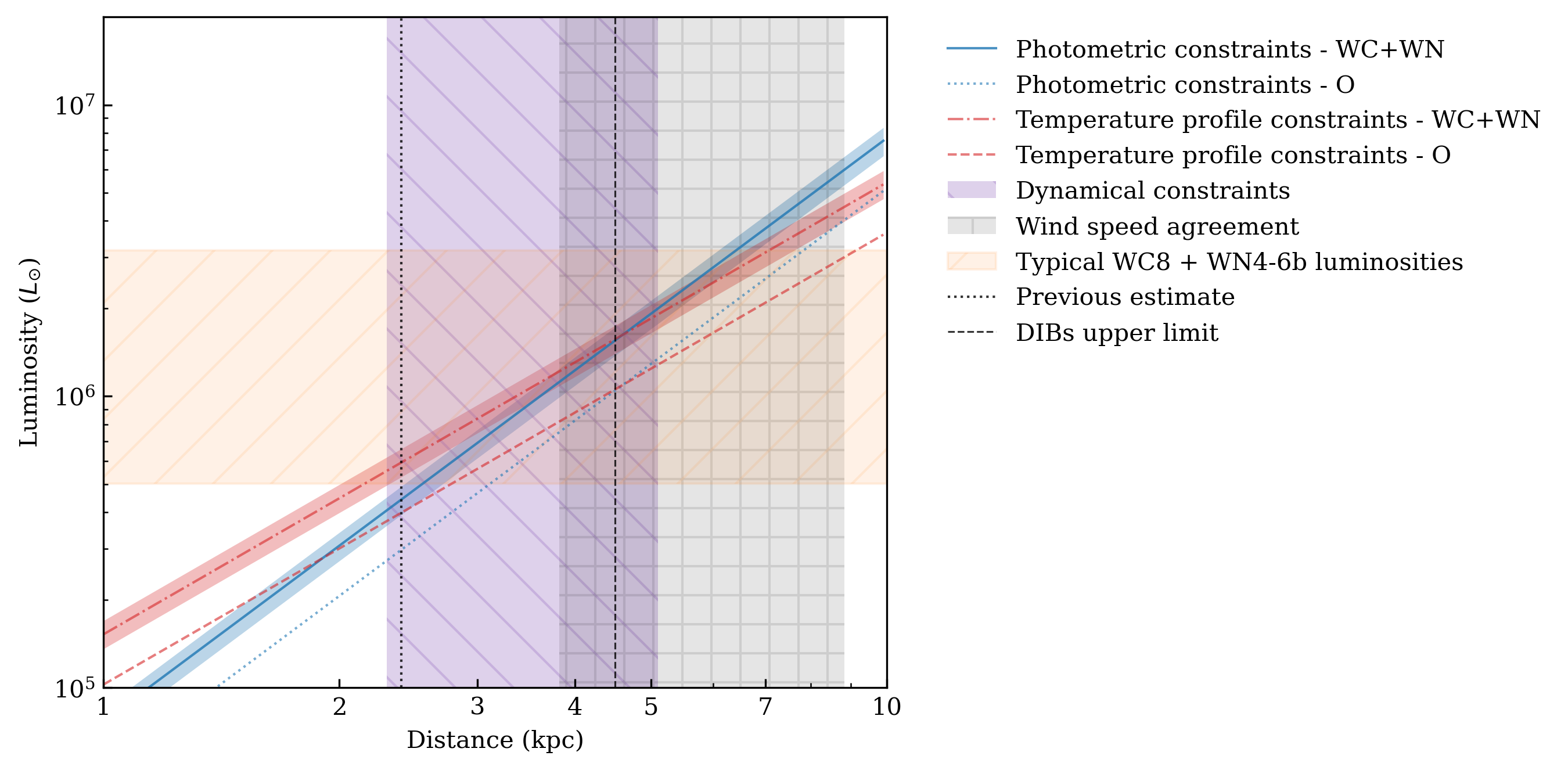}
    \caption{Plausible regions within the luminosity and distance parameter space for Apep based on constraints from its temperature profile, near-IR flux density, likely luminosity range and likely dynamical mass range given the orbit. The distances required for Apep to be explained by a spherically symmetric wind outflow within a factor of 2 is shown for comparison. }
    \label{fig:distance}
\end{figure*}

Our radiative transfer simulations suggest that unlike blackbody grains for which temperature scales with luminosity as $L^{0.25}$, the temperature instead scales with $L^{0.19}$, which more closely matches the analytical prediction based on the $1/\lambda$ approximation of the absorption coefficient as derived in Section~\ref{sec:temperature_profile}. 
To maintain the temperature profile measured at any given angular radius, $r/d$, $L$ must therefore be proportional to $d^{1.63}$, were $d$ is the distance of Apep, as shown in Fig.~\ref{fig:distance}. 

Importantly, this is different from the dependence between $L$ and $d$ provided by flux density constraints, in which case $L$ scales with $d^2$ to maintain a given flux density for the system. 
\citet{Callingham2019} used the V-band apparent magnitude to place constrains on the distance based on typical WR absolute magnitudes. 
Here, we use the near-IR SED \citep{Han2020} of the WR binary and the northern companion in the system to calibrate the scaling between luminosity and distance. 
Using the same WC:WN:O luminosity ratio and Potsdam WR and O star spectra as assumed in Section~\ref{sec:luminosity} to maintain their near-IR flux density ratio, the required scaling between the luminosity of the WR binary and the O star companion is shown in Fig.~\ref{fig:distance}. 

The different dependence between luminosity and distance required under the thermal profile and photometric constraints demands that the two constraints cross at only one point (apart from the origin) within the luminosity and distance parameter space, which in this case is 4.6\,kpc. This is almost twice the value from previous estimates \citep{Callingham2019, Han2020}. 
In practice, uncertainties in both constraints relax the point of intersection into a region of parameter space over a wide interval of plausible distances. However, further distance constraints are provided by the dynamics the binary's orbit. Given the orbital parameters derived for the system based on the dust geometry \citep{White2025}, the present position angle of the WR binary determined with near-IR interferometry \citep{Han2020} and radio imaging of the colliding-wind region \citep{Marcote2021} places it near apastron, translating the $47 \pm 6$\,mas projected separation \citep{Han2020} to a semi-major axis of $28_{-6}^{+10}$\,mas. Given the 193\,yr orbital period that we find, only a limited range of distances can permit such a semi-major axis for typical WC+WN masses between 20 and 40\,M$_\odot$ \citep{Crowther2007, Sander2019, Han2020}, which is indicated in Fig.~\ref{fig:distance}. The intersection between the thermal, photometric and dynamical constraints limit the range of plausible distances to between 2.8 to 5.1\,kpc, which also sits comfortably within the typical range of combined WN and WC luminosities given the spectral classification of Apep \citep{Hamann2006}, with the northern companion also at luminosities comparable to other stars of its spectral type \citep{Martins2005, Crowther2009}. Note that there could be additional lines of evidence related to Apep's distance that are not considered in depth here. These include diffuse interstellar bands (DIBs) present in the spectra of the system, which suggest an upper limit of 4.5\,kpc \citep{Zasowski2014, Callingham2019}; the radio luminosity of Apep, which would be similar to the baseline level (i.e., not during outburst) of the well-studied colliding-wind binary $\eta$\,Car \citep{Davidson1997} at this upper limit from DIBs \citep{Callingham2019}; and the X-ray luminosity  of Apep, which would be among the most luminous colling-wind binary known at this distance \citep{Callingham2019}.

In Fig.~\ref{fig:distance}, we also show the distance range required for the sky-plane dust outflow speed to be between 0.5 and 1 times the faster $3500 \pm 100$\,km\,s$^{-1}$ WN wind speed. 
his region of parameter space intersects with the intersection of the thermal, photometric and dynamical constraints between 3.8 and 5.1\,kpc. In order for the dust expansion rate and wind speed to be equal, a further distance of 8.2\,kpc is required. While this is beyond the most plausible range of distances based on the intersection from the constraints available, it is not forbidden by the thermal, photometric and dynamical constraints given their uncertainties. While this larger distance estimate does not fully resolve the line-of-sight and sky-plane outflow speed discrepancy, the degree of disagreement may not be as significant as previously thought, and any wind speed anisotropy invoked would only be required to account for a wind speed asymmetry by a factor of 2. Future observations by ALMA may be able to detect and characterise any asymmetric wind velocity field.

\section{Conclusions}
\label{sec:conclusions}
 
We summarise the main findings of this study below. 

\begin{enumerate}
    \item We imaged the colliding-wind Wolf-Rayet binary Apep with JWST/MIRI and VLT/VISIR. The JWST images detected 4 concentric dust shells with highly regular and detailed structures surrounding Apep. The mean expansion speed of the dust shells is $90 \pm 4$\,mas\,yr$^{-1}$ and the mean spacing between neighbouring shells is $17.30 \pm 0.17$\,arcsec. The shell spacing and expansion speed together suggest an orbital period of $193\pm11$\,yr, which is independent of uncertainties on the distance, and that the dust structure observed was produced over the past 700\,yr. 

    \item The expansion speed and shell spacing of the dust structure is subtly asymmetric across azimuth, possibly due to orbital modulation of the wind-shock conditions, orbital motion or a non-spherical stellar wind field.

    \item We determined the temperature of dust as a function of radius by modelling the spatially resolved SED of the dust structure assembled from the multi-wavelength observations. 
    Assuming amorphous carbon grains, the temperature profile that we measured is largely consistent with thermal equilibrium. 

    \item Combining thermal, photometric and dynamical constraints, we find that the likely distance to Apep is $4.6_{-1.8}^{+0.5}$\,kpc. A distance of 4.6\,kpc would imply a WR binary (combined) luminosity of $(1.6 \pm 0.2) \times10^6\,\mathrm{L}_\odot$ and an O star companion luminosity $(1.1 \pm 0.1)\times10^6\,\mathrm{L}_\odot$. Such a distance estimate does not fully remove the difference between the line-of-sight and sky-plane wind speed measured, but implies that their difference by a factor of $\sim$2 is smaller than previously estimated. 

\end{enumerate}

Using an example from a rare class of colliding-wind WR binaries, this study demonstrates the robust constraints that can be derived from resolved imaging on the dynamical, thermal and wind conditions of dust formation which has remained elusive. Future studies are encouraged to further explore the wind and dust dynamics and the thermal evolution of dust in a broader sample of dust-producing WR binaries.

\begin{acknowledgments}
This work is based on observations made with the NASA/ESA/CSA \textit{James Webb Space Telescope}. The data were obtained from the Mikulski Archive for Space Telescopes at the Space Telescope Science Institute, which is operated by the Association of Universities for Research in Astronomy, Inc., under NASA contract NAS 5-03127 for JWST. These observations are associated with program \#5842. Support for program \#5842 was provided by NASA through a grant from the Space Telescope Science Institute.
YH is funded by a Caltech Barr Fellowship. 
RMTW acknowledges the financial support of the Andy Thomas Space Foundation. 
BJSP was funded by the Australian Government through the Australian Research Council DECRA fellowship DE210101639. 
NDR is grateful for support from the Cottrell Scholar Award \#CS-CSA-2023-143 sponsored by the Research Corporation for Science Advancement.
\end{acknowledgments}

\begin{contribution}

YH planned the JWST and VLT observations, carried out data reduction and analyses and wrote the manuscript. All authors contributed to discussing scientific interpretations and editing the manuscript. 


\end{contribution}

\section*{Data availability}
The JWST/MIRI data used in this study is available on the Mikulski Archive for Space Telescopes under JWST program ID 5842 or at
\dataset[doi:10.17909/mf14-ga19]{https://dx.doi.org/10.17909/mf14-ga19}. The VLT/VISIR data used in this study are available on the ESO archive under program IDs 097.C-0679, 0101.C-0726 and 113.26A7.001.

%
\facilities{JWST(MIRI), VLT(VISIR)}

\software{\texttt{NumPy} \citep{numpy}, 
          \texttt{SciPy} \citep{scipy},
          \texttt{Matplotlib} \citep{matplotlib},
          \texttt{Astropy} \citep{astropy:2022},
          JWST Science Calibration Pipeline \citep{jwst_pipeline}
          }


\appendix

\section{Shell displacement fitting}
\begin{figure*}
    \centering
    \includegraphics[width=\linewidth]{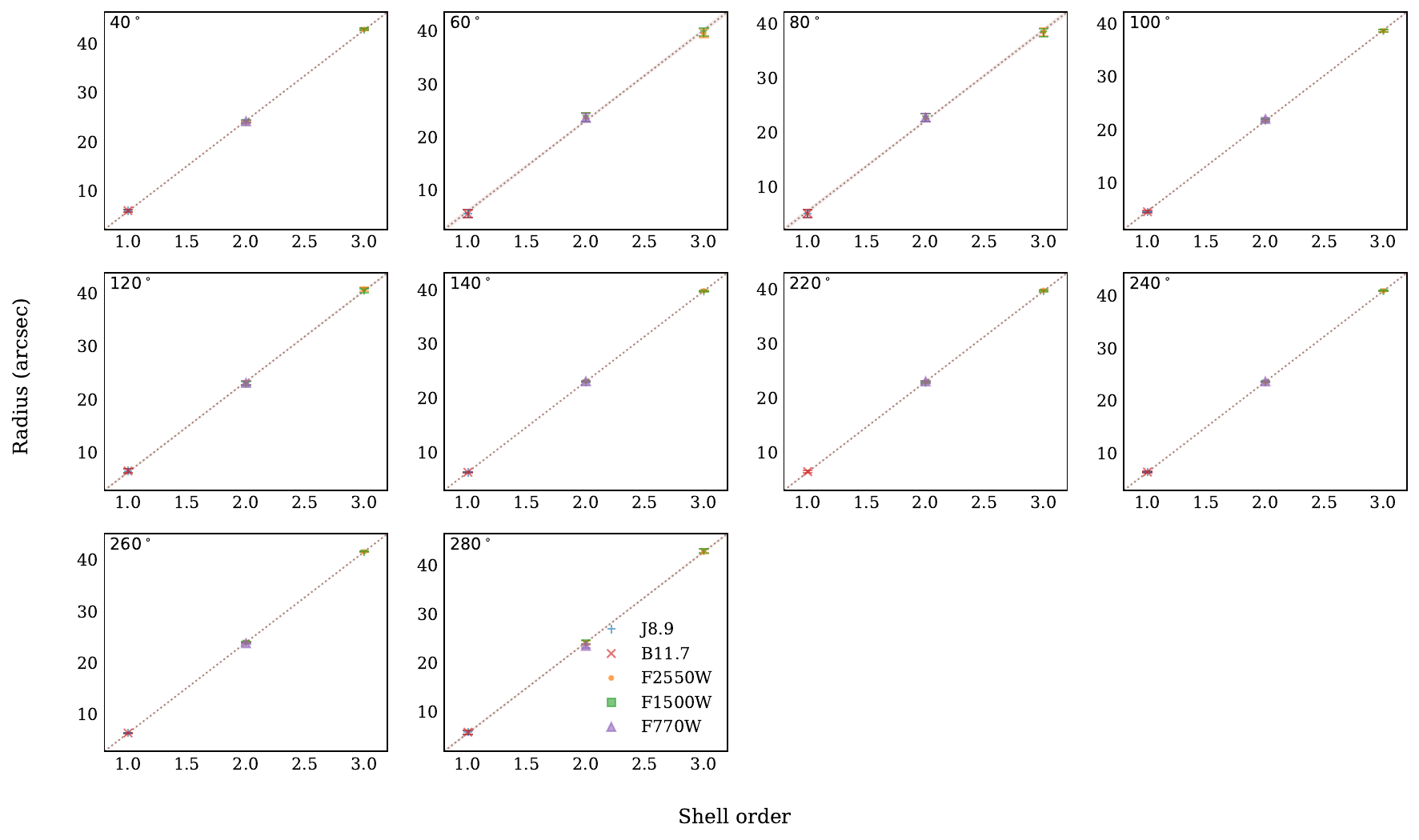}
    \caption{Linear fits to the shell radius as a function of shell order used to determine the inter-shell spacing along different position angles from the binary. }
    \label{fig:spacing_fits}
\end{figure*}

\begin{figure*}
    \centering
    \includegraphics[width=\linewidth]{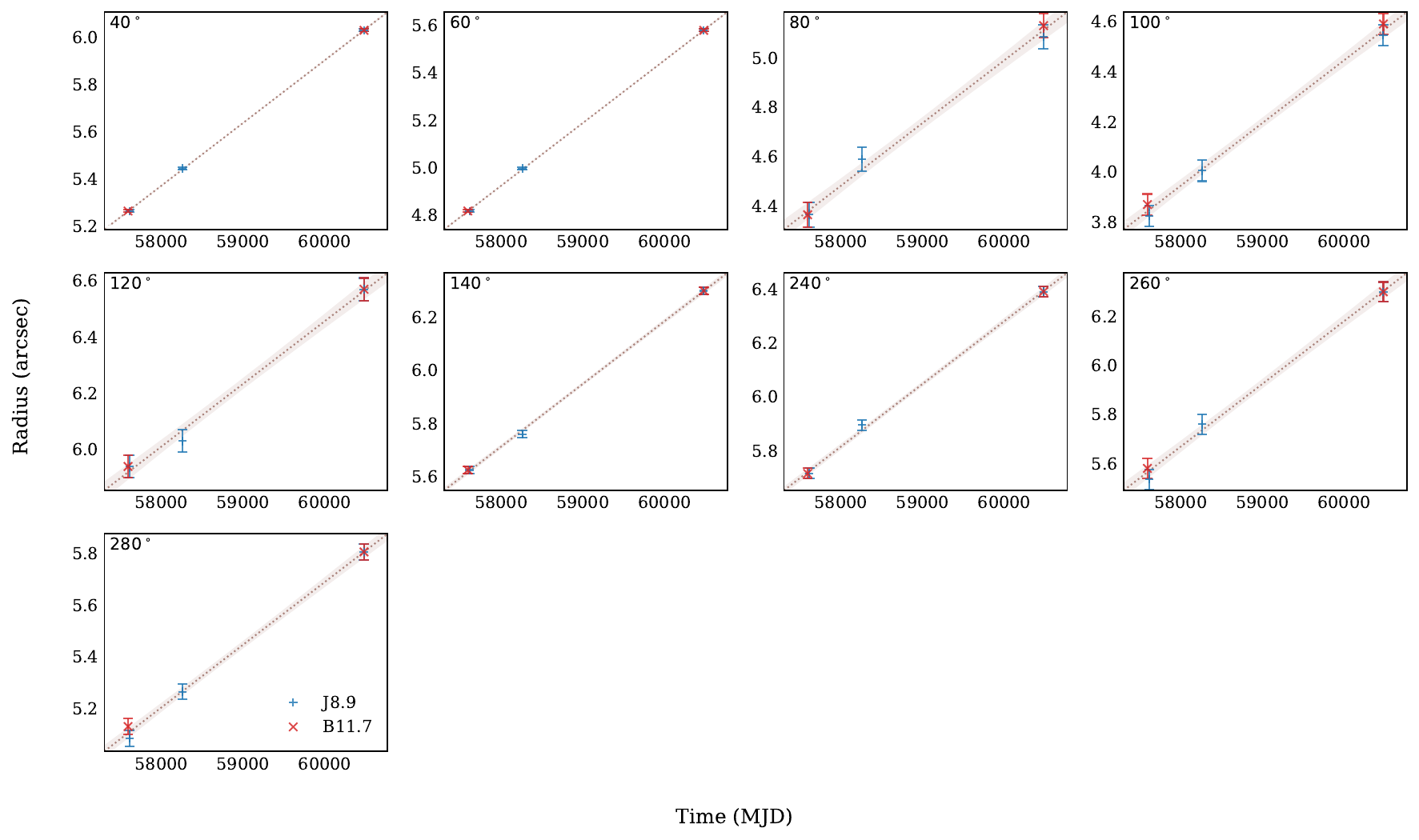}
    \caption{Linear fits to the inner shell radius as a function of time used to determine the expansion speed in different directions from the binary. }
    \label{fig:pm_fits}
\end{figure*}

The displacement and expansion speed fitting of dust shell edges are shown in Fig.~\ref{fig:spacing_fits} and Fig.~\ref{fig:pm_fits}. 

\section{Resolved SED fitting}
\begin{figure*}
    \centering
    \includegraphics[width=0.7\linewidth]{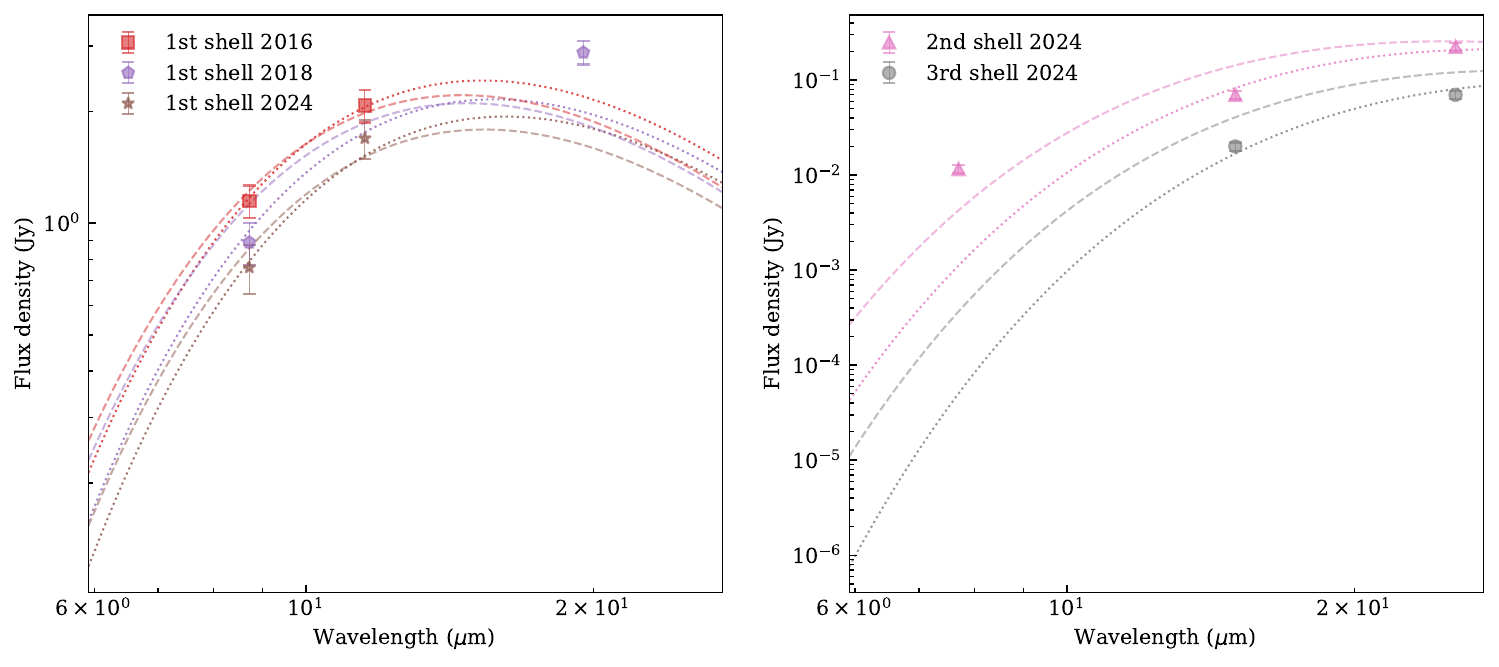}
    \includegraphics[width=0.7\linewidth]{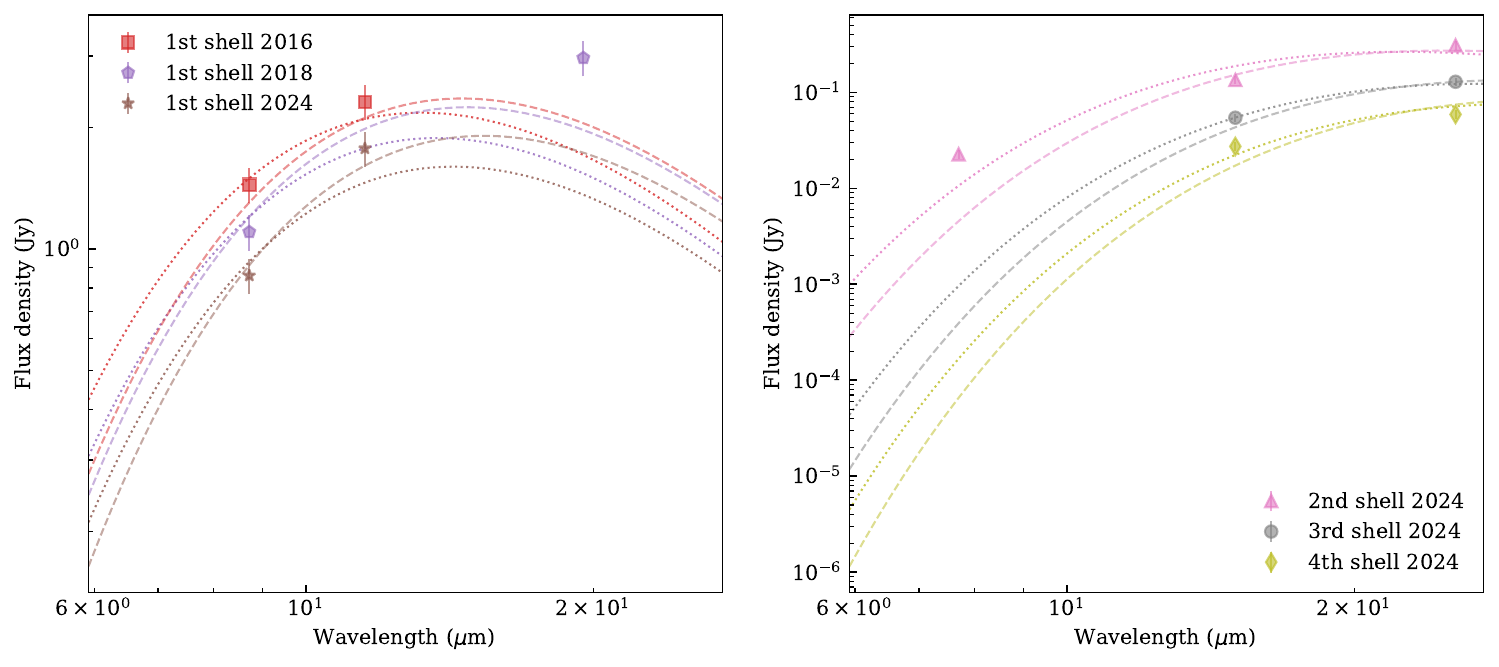}
    \includegraphics[width=0.7\linewidth]{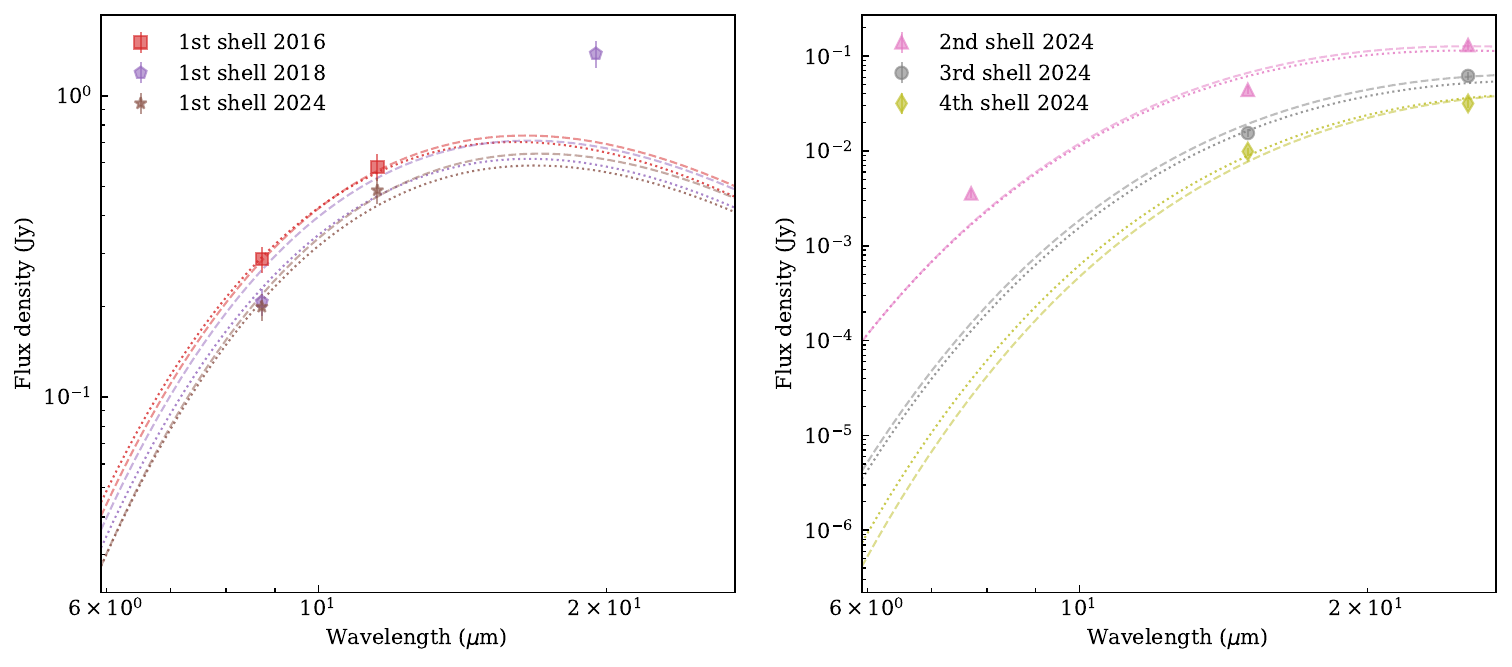}
    \caption{The spatially resolved SED of Apep constructed from a range of shell orders and observing dates. The top, middle and bottom rows correspond to the E1, E2 and E3 features respectively. The SED models fitted individually to each feature at each epoch (Section~\ref{sec:sed}) are shown with dotted lines, whereas the SED models assuming the best-fit power-law temperature profile (Eq.~\ref{eq:tpl}) are shown with dashed lines.} 
    \label{fig:sed_fits}
\end{figure*}

The resolved SEDs are shown in Fig.~\ref{fig:sed_fits}.

\bibliography{references}{}
\bibliographystyle{aasjournalv7}



\end{document}